# Adaptation des réseaux sociaux d'entreprise pour favoriser l'apprentissage informel sur le lieu de travail

Carine Touré[1,3], Christine Michel[1], Jean-Charles Marty[2]

[1] INSA de Lyon, Univ Lyon, CNRS, LIRIS, UMR 5205, F-69621 Villeurbanne, France,
[2] Université de Savoie Mont-Blanc, CNRS, LIRIS, UMR 5205, F-74000 Chambéry, France,
[3] Société du Canal de Provence, Le Tolonet, France
{Carine-Edith.Toure, Christine.Michel, Jean-Charles.Marty}@liris.cnrs.fr

**Résumé.** L'apprentissage informel sur le lieu de travail est réalisé lors des activités quotidiennes auxquelles participent les collaborateurs et représente plus de 75% de l'apprentissage en entreprise. Les réseaux sociaux d'entreprise sont massivement utilisés actuellement pour promouvoir ce type d'apprentissage. A partir d'une étude de terrain, nous montrons qu'ils sont effectivement adaptés concernant les aspects sociaux mais que la conception doit être repensée pour satisfaire les besoins contextuels des utilisateurs concernant le contenu et l'accès au corpus informationnel, les indicateurs de qualité de l'information et les formes de modération et contrôle.

**Mots-clés.** Apprentissage informel, apprentissage tout au long de la vie, réseau social d'entreprise, conception centrée utilisateur, apprenant adulte

**Abstract**. The informal learning in the workplace is realized during daily collaborators' activities and represent more than 75 % of the learning occurring in a company. Enterprise social networks are currently massively used to promote this type of learning. From a pilot study, we show that they are actually adapted concerning the social aspects, but that the design must be rethought to consider user contextual needs linked to the content and access to informational corpus, the information quality indicators and the forms of moderation and control.

## 1 Introduction

L'apprentissage tout au long de la vie est considéré depuis les années 1970 comme une approche de l'éducation qui permet de construire des connaissances et compétences nécessaires pour réussir dans un monde en évolution rapide [1]. Il se décline en : apprentissage formel, non formel et informel [2]. Les apprentissages formel et non formel ont la caractéristique d'être structurés par outils ou séquences de formation à l'inverse de l'apprentissage informel. Ce dernier se produit au cours des expériences quotidiennes, en travaillant ou en interagissant avec d'autres personnes. Il correspond aux apprentissages faits lors des activités de travail quotidiennes auxquelles participent les collaborateurs [3] et est motivé par un besoin personnel. L'apprentissage informel revêt une importance capitale pour les entreprises, car il représente plus de 75% de l'apprentissage au travail [4]. C'est le moyen le plus important d'acquérir et de développer les connaissances et compétences requises en contexte professionnel. Des recherches en gestion des connaissances (KM) ont étudié comment promouvoir la gestion et le maintien du partage des connaissances sur le lieu de travail. Pour l'apprentissage informel, trois générations de technologie ont été privilégiées [5] [6]. Elles s'articulent autour des deux grandes catégories de stratégies proposées pour gérer les connaissances : valoriser le capital informationnel et valoriser le capital humain par la collaboration [5] [7].

La première génération partait du principe que le collaborateur peut se former en continu et identifier comment résoudre les problèmes qu'il peut rencontrer sur son lieu de travail en cherchant directement de l'information sur les procédures et savoir-faire liés à son activité. Pour soutenir cette démarche, les entreprises ont produit des corpus d'information relativement exhaustifs sur l'activité professionnelle et les pratiques développées au cours du temps, et les ont rendus accessibles aux collaborateurs. Les bases de connaissances produites, pourtant très riches, sont restées souvent inutilisées car elles étaient inadaptées aux besoins et caractéristiques des collaborateurs en particulier en termes d'accès à l'information et de formation [8] [9]. De plus, les fonctionnalités des outils d'accès à cette information ne sont pas dédiées au processus d'apprentissage. Sur ce dernier point, Grasser [9] recommande de privilégier des objectifs de formation basés sur l'auto-régulation et la méta-cognition de manière à former les apprenants à apprendre à apprendre et décrit [10] différents outils et principes opérationnels basés sur le plaisir, la rétro-action ou le contrôle pour les soutenir.

La deuxième génération s'est axée sur le partage d'expertise et l'identification des personnes ressources les plus à même de fournir l'information utile au collaborateur cherchant à se former. La démarche de structuration de Communauté de Pratiques (CoP) a été la plus communément adoptée par les entreprises pour aider les praticiens à exprimer et partager leurs pratiques, et à développer et exploiter leurs connaissances [11] [7]. L'interaction directe entre pairs a été reconnue pour faciliter le transfert des connaissances et améliorer la qualité de l'information reçue [12] mais a montré ses limites en termes d'exhaustivité de l'information, de précision dans l'identification et la recommandation d'experts à partir de leurs profils ou de garantie de la protection et du contrôle de leur vie privée [5]. Les CoP sont restées de plus souvent peu utilisées.

La troisième génération a cherché à fusionner les principes des outils de première et seconde génération et est caractérisée par l'exploitation d'espaces collaboratifs d'information qui combinent le dépôt d'information, la communication et la collaboration. De nombreuses entreprises ont fait le choix de s'orienter vers l'exploitation des Réseaux Sociaux d'Entreprise (RSE) pour améliorer les performances organisationnelles, en particulier dans le contexte du partage des connaissances [13]. Ils intègrent sur une même plateforme la gestion de l'activité professionnelle et la stratégie de gestion des connaissances et du savoir-faire et les aspects collaboratifs et sociaux favorisant l'interactivité entre pairs [14] [15] [16]. Les RSE offrent l'avantage de valoriser le capital informationnel et social et sont particulièrement adaptés pour trouver et interagir avec d'autres collaborateurs, demander de l'aide et y répondre [5]. Ils sont plus faciles d'utilisation, plus attractifs et interactifs que les environnements de collaboration traditionnels et répondent aux besoins d'utilité et de gratification des utilisateurs [17]. En effet, outre la facilité pour communiquer et partager de l'information, ils permettent la reconnaissance de chacun dans les contributions faites et une forme de reliance sociale matérialisée par le simple fait de suivre un post ou de le commenter. Néanmoins, l'exploitation ouverte des multiples fonctionnalités d'accès à l'information, de contribution et collaboration ont montré des utilisations abusives conduisant à un manque d'efficacité dans l'exploitation des ressources informationnelles ou un sentiment de harcèlement [18].

L'objectif de cet article est d'étudier dans quelle mesure les RSE sont effectivement de bons outils pour mettre en œuvre des stratégies d'apprentissage informel. Plus spécifiquement, nous étudierons quelles fonctionnalités des RSE sont les plus efficaces pour atteindre les objectifs d'apprentissage recommandés par Grasser et comment les adapter pour qu'elles soient cohérentes avec les objectifs et pratiques de l'organisation industrielle et des collaborateurs pour en favoriser un usage durable et pérenne. Pour répondre à ces questions nous avons, dans la section suivante, étudié les caractéristiques des RSE et comment ils peuvent soutenir de manière significative l'apprentissage informel en contexte professionnel de manière à identifier différentes propositions de conception. Nous avons testé ces propositions dans un contexte industriel pour en évaluer la pertinence et les raffiner. Cette étude est présentée dans la troisième partie de cet article.

## 2  Utilisation des RSE pour l'apprentissage informel

**Avantages.** Les fonctionnalités des RSE favorisent la *construction et l'identification d'informations pertinentes*. Les commentaires au sein des médias sociaux sont une forme d'expression et de communication emblématique qui donne aux utilisateurs le sentiment de pouvoir juger efficacement de la qualité d'une information et de participer facilement à la construction de contenus. En effet, l'obsolescence de l'information proposée dans les outils d'apprentissage informels est un problème récurrent. Les commentaires ont cet avantage que les collaborateurs peuvent communiquer et participer, en ligne et juste à temps, à la construction du corpus de connaissances [19], ce qui atténue le risque d'oubli ou de perte de pratique. Les appréciations laissées par les lecteurs jouent également un rôle sur le jugement de l'information publiée mais aussi sur la motivation à contribuer. Elles peuvent être plus formalisées comme c'est le cas de certains wikis qui présentent des indicateurs de complétude ou de lisibilité des articles. Ces indicateurs permettent aussi à un collaborateur de juger rapidement l'article et de mieux comprendre comment il peut participer à son amélioration. Cette rétroaction rapide matérialise pour un contributeur l'utilité concrète de l'information publiée [20] et participe à construire sa réputation. Certains outils comme les wikis exploitent d'ailleurs souvent ces fonctionnalités pour soutenir le travail collaboratif d'innovation, de résolution de problème et plus globalement pour aider les organisations à améliorer leurs processus d'affaires [18].

Les médias sociaux donnent de plus de la *visibilité et de la persistance à de multiples actions communicatives* comme le téléchargement et la publication de contenu et l'identification de qui a communiqué sur quoi, mais aussi la mise à jour de son statut, la définition de son profil et la mise en valeur d'aspect particulier de soi-même ou la connexion avec d'autres personnes [15] [21]. Ils élargissent ainsi (et précisent) la gamme de personnes, de réseaux et de textes à partir desquels les gens peuvent apprendre à travers l'organisation.

La révélation des comportements des uns et des autres via les notifications, nombre de messages, nouvelles soumissions, ..., peut aussi être utilisée pour *s'auto-réguler* en identifiant les pratiques des autres, en évaluant les siennes et en ajustant son propre comportement, *favorisant ainsi les processus de méta-cognition et la capacité à apprendre à apprendre* [22]. Cette prise de conscience peut aussi être source de motivation pour construire sa propre identité numérique au travers de ces indicateurs [23]. L'inscription dans un groupe permet enfin aux collaborateurs de développer des connaissances méta-sociales et de faciliter leur capacité à se coordonner et à collaborer [24], en particulier au sein des CoP.

**Limites.** [18] identifie deux groupes de risques : ceux liés à l'acceptation et à la capacité d'utilisation des médias sociaux et ceux liés aux contenus publiés par les collaborateurs.

L'acceptation et la capacité d'utilisation jouent un rôle fondamental sur l'utilisation initiale et continue de ces technologies. L'acceptation s'amorce par la construction de premières croyances à propos du système d'information, générées par des stimuli externes tels que la qualité du système, la qualité du service, la qualité des connaissances ou la qualité de l'information [25] [26] [27] [28]. Ces croyances sont modérées par des facteurs propres à l'utilisateur comme son âge, son expérience préalable ou par l'accompagnement qu'on lui propose [29]. Ils conditionnent aussi la capacité d'utilisation. En effet, une utilisation efficace des RSE nécessite un niveau élevé d'alphabétisation et de maîtrise technique des systèmes numériques pour rechercher et extraire de l'information ou pour porter un jugement critique sur sa véracité et son utilité et ainsi trouver l'information utile ; ou pour interagir avec les ordinateurs ou les personnes à distance [30] [18]. Les caractéristiques contextuelles des collaborateurs sur le lieu de travail ne sont généralement pas considérées dans la conception [3]. Pour développer des compétences méta-sociales et améliorer leur communication tel que [10] le mentionne, les utilisateurs doivent ressentir la réciprocité de leurs pairs. Ils ont besoin de clarté sur les objectifs, d'être conscients de la qualité de l'information partagée. Ils ont également besoin d'avantages intrinsèques de leurs actions (amélioration de la réputation professionnelle, être mieux connus dans la communauté, être informés que leurs actions ont été appréciées par les autres) [12]. De plus, des formes d'accompagnement et de formation ou des formes de gouvernance (contrôle, surveillance et filtrage) des accès doivent être intégrées [18]. L'analyse de ce type de besoins est coûteuse et les entreprises qui cherchent des solutions rapides et légères ne les financent souvent pas. Ce premier cycle d'utilisation construit chez l'utilisateur une nouvelle expérience et de nouvelles croyances confirmant ou non les précédentes, ce qui modifie son attitude (satisfaction ou insatisfaction) et son intention d'utiliser le système sur le long terme [31] [32].

Le deuxième groupe de risques concerne la validité et la qualité des informations créées et publiées. Bien que l'information publiée ne soit pas anonyme, elle peut se révéler être inutile pour servir de base à un apprentissage informel efficace pour des raisons de manque de détail et de précision, en particulier sur le plan technique. Les publications des RSE sont en général assez courtes, très spécifiques ou au contraire trop générales. Elles sont adaptées aux mises à jour, mais pas à la construction du corpus d'information de base. En outre, les messages postés peuvent être inappropriés et donner aux collaborateurs un sentiment de harcèlement. Les notes et les commentaires sont souvent considérés comme des jugements. La capacité de discerner le type d'information partageable est pour la plupart du temps imputable aux utilisateurs et ils sont peu modérés ou contrôlés, c'est le principe même d'un média social [30] [18]. Ces risques peuvent affecter négativement l'environnement social et remettre en cause le processus d'apprentissage informel recherché.

**Bilan et proposition.** Les RSE semblent donc être de bons candidats pour soutenir l'apprentissage informel en entreprise. Ils fournissent des fonctionnalités qui favorisent et facilitent la collaboration, le partage des connaissances, la visibilité des utilisateurs, la valorisation et la persistance de l'information. Ils proposent également des indicateurs réflexifs qui facilitent l'analyse et la coordination des activités collectives, du lien social et de l'apprentissage. Ces caractéristiques placent les collaborateurs et leurs besoins au cœur de l'environnement d'apprentissage, ce qui rend ces outils appropriés pour soutenir l'apprentissage informel sur le lieu de travail. Néanmoins, leur utilisation peut se révéler inefficace car des apprenants de type « travailleurs adultes » doivent être conscients de la valeur de leur participation au groupe d'apprentissage. Ils recherchent une rétroaction personnelle et professionnelle, une forme d'utilité et gratifications concrètes. De plus, la question de la qualité de l'information publiée peut poser problème pour des stratégies d'apprentissage.

Pour pallier les risques liés à la qualité de l'information, il nous semble fondamental de baser la conception de l'environnement sur *un corpus informationnel précis et relativement exhaustif des procédures et savoir-faire déjà formalisés dans l'entreprise en l'enrichissant des caractéristiques des espaces collaboratifs*. La revue de littérature a montré que l'indexation et la structuration de l'information devaient être repensées pour en faciliter l'accès en contexte. Un moteur de recherche en langage naturel et des tags d'indexation par mots clés sont des éléments fondamentaux pour garantir un accès transversal à l'information. L'aspect contextuel lié à l'activité tel que le proposent les communautés de pratique peut être reproduit par des wikis structurés selon les communautés de travail de l'entreprise. Différents éléments sont à considérer pour garantir d'une part la qualité et la confiance en l'information publiée et d'autre part faciliter la contribution : sélectionner l'information juste utile à la communauté, organiser l'information selon les modèles spécifiques de documents liés à cette activité et organiser la validation selon les structures hiérarchiques de décision de la communauté.

Concernant le soutien à l'apprentissage, la revue de la littérature a mis en évidence trois caractéristiques complémentaires des médias sociaux pour promouvoir et faciliter la valorisation des utilisateurs, la visibilité et la réflexivité. Les *commentaires et appréciation*s (comme les « like ») peuvent être considérés comme des outils d'expression et de communication, permettant aux collaborateurs de fournir des éléments de rétroaction et de participer à la construction des contenus. Ces fonctionnalités leur permettent d'être impliqués dans la co-construction des connaissances et de maintenir à jour les informations disponibles, éléments importants pour la qualité des processus d'apprentissage. Les *indicateurs d'awareness* de type notifications (nouvelles contributions, qui et quand, nombre de commentaires) peuvent favoriser la construction des compétences métacognitives d'auto-régulation et stimuler la participation. Les *indicateurs sur la qualité de l'information* peuvent faciliter l'identification de l'information utile et la collaboration par une analyse critique des éléments complémentaires à renseigner pour mettre à jour et améliorer les contenus.

Enfin, pour limiter les risques de non utilisation de l'environnement, nous proposons d'utiliser une *démarche incrémentale et itérative de conception centrée utilisateur*. Cette démarche offre l'avantage d'identifier les caractéristiques et préférences des utilisateurs et de concevoir un environnement adapté à leur contexte professionnel. Le caractère incrémental et itératif de la démarche permet aussi d'accompagner le changement lié à l'introduction d'un nouveau système d'information et ainsi d'influer positivement sur son acceptation et son utilisation initiale et continue. En effet, l'apprentissage informel relevant par nature de la volonté du collaborateur et n'étant pas stimulé par des stratégies d'accompagnement, cette caractéristique est fondamentale. L'analyse des modèles de l'acceptation technologique des SI sur le lieu de travail a montré que le modèle de l'acceptation peut être représenté par une spirale (voir figure 1) structurée par les modalités d'utilisation. Chaque cycle construit un artefact technologique de plus en plus adapté aux besoins et usages de l'utilisateur. Nous faisons l'hypothèse qu'il est possible, en concevant des artefacts progressivement de plus en plus adaptés aux besoins et croyances des utilisateurs, de favoriser l'utilisation durable et pérenne nécessaire à l'apprentissage informel.

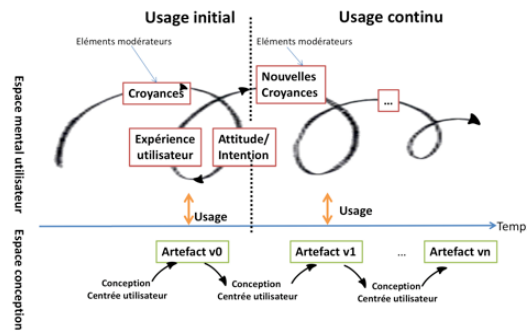

**Fig. 1.** Conception incrémentale et itérative des SI pour l'apprentissage informel

Ces différents principes ont été testés dans un contexte réel industriel. L'objectif était d'une part d'identifier les éléments de conception des médias sociaux les plus adaptés pour soutenir l'apprentissage informel, tester la mise en œuvre de notre méthodologie de conception incrémentale et itérative et dégager un ordre structurant les éléments de conception à considérer dans chaque cycle. Nous présentons les résultats de cette expérimentation dans le chapitre suivant.

## 3   Mise en œuvre de la méthodologie

### 3.1   Contexte et constitution du groupe de travail

La Société du Canal de Provence (SCP) est spécialisée dans les services liés au traitement et à la distribution de l'eau pour les entreprises, les agriculteurs et les communautés. Son territoire d'intervention est divisé en dix zones géographiques appelées Centres Opérationnels (CO). Chaque CO correspond à une communauté de pratique qui regroupe 3 types de collaborateurs : l'Opérateur (O), le Technicien Coordinateur (TC) (un opérateur qui a également le rôle de gestionnaire de la communauté) et le Technicien Support et Relation Client (TSRC). Ils sont responsables de l'entretien des infrastructures hydrauliques (canaux, stations de pompage, stations d'épuration des eaux, etc.). Les collaborateurs doivent maitriser de nombreuses procédures et informations qui sont amenées à changer régulièrement du fait de l'évolutivité du réseau et des normes d'exploitation. Pour les assister, la SCP a produit en 1996 un livre de connaissances au format HTML sur les procédés et les infrastructures hydrauliques : ALEX (Aide à l'exploitation). En 20 ans, Alex n'a que très peu été utilisé principalement car les moyens proposés pour accéder à l'information étaient inadaptés, le principe d'avoir une plateforme d'assistance étant en revanche soutenu par les collaborateurs. ALEX est en ce sens un exemple emblématique des stratégies de KM basées sur les livres de connaissances produits dans les années 1990-2000 et un bon contexte pour travailler sur les moyens les plus à même de soutenir l'apprentissage tout au long de la vie.

Quatre CO ont été sélectionnés par le responsable du projet pour agir comme pilotes dans cette étude. Onze employés distribués sur ces centres ont été choisis de manière à couvrir différents métiers de l'entreprise et avoir des utilisateurs néophytes et expérimentés d'Alex. Ils ont été invités par la direction à participer librement au groupe de travail. Les groupes de discussion ont été animés par les auteurs et modérés par un des membres du CA responsable du projet ALEX. Au total 12 séances de travail ont été réalisées sur une période de 2 ans. La première année a consisté à formaliser les besoins fondamentaux. Six réunions, espacées de 2 à 3 semaines, ont permis d'affiner une solution sur la base de prototypes de plus en plus évolués jusqu'à avoir une solution opérationnellement utilisable en contexte de travail. La solution a été développée et mise en exploitation pendant 3 mois. A l'issue de cette première année un bilan a été fait sur la recevabilité de la solution proposée et un nouveau cycle d'analyse et de co-conception a été amorcé. Il s'est déroulé sur sept mois. Six séances de travail ont été réalisées.

### 3.2   Eléments structurant le premier cycle de conception

Les résultats de la première boucle sont présentés plus en détail dans [33]. En synthèse, cette étape a montré que la principale exigence était de proposer des moyens plus faciles pour rechercher, soumettre et accéder aux connaissances (cf Figure 2 zones 1,5,7) en les organisant sous la forme d'une communauté de pratique structurée par CO selon des espaces collaboratifs. Les discussions ont permis d'identifier la structure générale de navigation et d'organisation de l'information du site et les modalités de structuration des connaissances, en particulier onze structures différentes de fiches d'expériences. Un travail d'harmonisation de l'architecture des différents SI a été réalisée pour intégrer Alex aux autres SI et à l'intranet de l'entreprise de manière à favoriser la navigation entre les différents outils et afin qu'ils soient facilement accessibles de chaque poste de travail et sur le terrain. Un espace simplifié reproduisant une suite bureautique de type tableur/traitement de texte et des modèles de documents ont été conçus pour produire les fiches. Quatre rôles ont été proposés pour contrôler les contributions et garantir la qualité de l'information - le *lecteur*, le *contributeur*, le *validateur*, le *gestionnaire*. Le groupe de travail a été responsable des affectations en tenant compte des CO et des métiers selon les niveaux de responsabilité existant. Par exemple la validation des fiches a été affectée, pour chaque centre, au responsable de centre et le rôle du gestionnaire a été attribué aux responsables du projet Alex.

Après 3 mois d'exploitation, une évaluation de l'outil a montré que le nouvel Alex remplissait les fonctionnalités centrales attendues par les utilisateurs mais manquait d'éléments attractifs pour en garantir l'usage sur le long terme [33]. La seconde étape du cycle de conception a permis de travailler sur ces éléments.

### 3.3     Eléments structurant le second cycle de conception

Les discussions se sont orientées vers la conception d'éléments permettant la stimulation, le contrôle et le suivi de l'activité. Elles ont fait émerger des besoins différents pour les lecteurs et contributeurs, et pour les validateurs et gestionnaires. Les premiers ont été sensibles à l'ajout de fonctionnalités sociales de valorisation/participation et d'indicateurs d'utilisation (commentaires, notations, notifications ...). Les seconds ont exprimé des attentes sur le fait d'avoir des indicateurs de suivi général du processus de construction de connaissances via un tableau de bord de supervision. Nous présentons dans la suite de ce chapitre les résultats liés aux fonctionnalités sociales, le tableau de bord n'étant pour l'instant pas achevé en termes de développement.

**Commentaires et appréciations.** Les discussions sur les commentaires et appréciations se sont basées sur des maquettes présentant les modalités d'interaction les plus couramment proposées dans les plateformes Web 2.0, les commentaires et les « like » comptabilisant le nombre d'appréciations positives. Tous les participants ont validé le principe des commentaires considérés comme plus légers pour la communication que les courriels et rendant également les feuilles plus interactives, le commentaire étant perçu comme un « outil d'annotation ». Les participants ont souligné la nécessité de notifier les contributeurs sur la présence de commentaires sur leurs fiches, ainsi que la modération et l'archivage des commentaires afin d'améliorer la lisibilité des fiches et de contrôler les excès potentiels ou le harcèlement des collègues. Les validateurs ont été désignés pour assurer la modération et l'archivage des commentaires lorsque les modifications suggérées étaient prises en compte dans les fiches. La fonctionnalité « like » a suscité de nombreuses discussions. Plusieurs participants ont exprimé un avis plutôt négatif et des inquiétudes liées aux abus potentiels (un « like » pouvant être attribué ou pas selon l'affinité que l'on peut avoir avec le contributeur et non sur la qualité de la contribution) et aux risques sur la motivation des contributeurs s'ils n'en reçoivent pas. D'autres participants, déjà familiers des réseaux sociaux, ont été enthousiastes et ont vu le côté « ludique » de la fonction. Suite aux discussions le « like » a été modifié en une appréciation plus impersonnelle - « feuille utile » - mais lors de la réunion suivante, les participants sont revenus sur leurs positions et ont finalement choisi d'utiliser la fonction « like » telle que. Ils y ont vu l'intérêt de permettre aux contributeurs d'avoir une rétroaction rapide sur leur contribution, une identification rapide des informations potentiellement intéressantes et le sentiment de faire partie d'une communauté de travail. Certaines adaptations ont été demandées : remplacer le pouce levé trop connoté par un émoticône souriant, et initier des discussions, entre les collaborateurs, sur cette fonctionnalité sociale pour prévenir les risques de malentendus et d'abus.

**Indicateurs d'activité.** Différents indicateurs réflexifs d'activité ont été discutés avec les participants : les notifications de commentaires posés, leur nombre par fiche, l'identification des auteurs/lecteurs avec leurs dates d'activité, le statut des contributions (commentaire ou fiche). Certains indicateurs comme le nombre de commentaires, appréciations et contributions n'ont déclenché aucune discussion dans la mesure où ces éléments avaient déjà été discutés dans les précédentes réunions (cf Figure 2 zone 3,4). Les notifications sur les nouvelles contributions publiées ou consultées ont été mentionnées pour faciliter l'identification de l'information récente et des centres d'intérêt des autres collaborateurs. L'identification des acteurs, comme le dernier contributeur ou le dernier lecteur a été jugée utile pour amorcer des discussions directes entre collègues. L'identification des contributeurs successifs n'a en revanche pas été jugée nécessaire, une fiche validée étant considérée comme un travail collectif. Le statut des publications (fiches en attente, refusées, acceptées) a émergé suite au besoin exprimé de savoir si et quand le validateur a pris en compte une contribution. Finalement, en se projetant dans des cas d'utilisation possibles, les discussions ont fait émerger deux modalités de présentation de ces indicateurs : dans une page personnelle liée à son profil (cf Figure 2 zone 6 pour un accès à cet espace) et dans la page d'accueil du CO. La première page a été vue comme un moyen proposé à chaque collaborateur de suivre sa propre activité et d'en voir la portée au sein de l'organisation. La seconde a été vue comme un moyen d'identifier la dynamique d'une communauté, l'information récente ou considérée comme utile et ainsi se former et amorcer les discussions entre collègues (cf Figure 3).

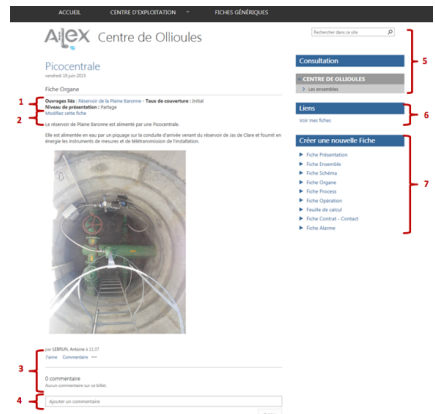

**Fig. 2.** Page de consultation d'une fiche

**Indicateurs de qualité de l'information.** Trois indicateurs ont été proposés pour exprimer la qualité de l'information : *lisibilité*, *exhaustivité* par rapport au concept décrit et *pertinence* [34]. L'objectif est d'informer l'utilisateur de l'effort de lecture nécessaire à réaliser pour mettre en pratique l'information présentée dans des situations réelles de travail ou des résolutions de problèmes. Les participants se sont globalement déclarés favorables à l'utilisation de ce type d'indicateurs. Les discussions ont porté sur les échelles d'évaluation, les modalités d'attribution des valeurs et le nom des indicateurs. Pour décrire la lisibilité d'une fiche, les participants ont proposé une échelle à 4 niveaux : *Opérationnel* (les informations sur la fiche sont immédiatement exploitables), *Assistance* (une procédure opérationnelle décrite mais exigeant une analyse pour l'adapter au contexte), *Acquisition* (informations générales décrivant un contexte) et *Partage* (ébauche de fiche nécessitant des contributions complémentaires). Le groupe a préféré le terme « Niveau de présentation » pour désigner cet indicateur. L'exhaustivité a été jugée utile en utilisant le nom « Taux de couverture » pour l'indicateur et une échelle à 3 niveaux : *faible*, *moyen*, *bon*. L'indicateur de pertinence n'a pas été jugé utile, une fiche n'étant publiée que si elle est considérée comme pertinente par le validateur et retirée si les commentaires la jugent inutile. Comme pour la validation des fiches, les participants ont décidé que c'était au validateur d'évaluer le niveau de présentation et le taux de couverture. Les participants n'ont pas jugé utile de représenter cette évaluation par une icône (étoiles, feux, ...) et ont préféré que les mentions soient directement écrites dans l'entête de chaque fiche (cf Figure 2 zone 1).

## 4 Discussions et conclusion

Notre analyse a montré que les RSE pouvaient être de bons candidats pour soutenir les stratégies d'apprentissage informel. Les fonctions sociales de type commentaires, appréciations, indicateurs d'activité sont en effet adaptées pour stimuler les usages et soutenir l'apprentissage, en particulier sur les aspects méta-cognitifs. Trois adaptations doivent être néanmoins réalisées : (1) baser la conception sur un corpus informationnel précis et relativement exhaustif des procédures et savoir-faire déjà formalisés dans l'entreprise et en contextualiser l'accès sous la forme de communauté de pratique structurée selon des espaces collaboratifs ; (2) ajouter des indicateurs de jugement sur la qualité opérationnelle de l'information et le capital informationnel construit, et (3) définir des formes de modération et contrôle cohérentes avec les structures hiérarchiques de l'entreprise.

L'étude a montré l'intérêt de mettre en œuvre, sur un temps long, une démarche incrémentale et itérative de conception centrée utilisateur pour définir, dans la discussion, comment adapter la conception et accompagner le changement. Deux cycles de conception ont été réalisés, un lié à la gestion de l'information, l'autre lié à la gestion de l'attention. Le fait de renforcer le travail de conception sur les architectures informationnelles, en termes de contenus, structuration et publication, n'est pas antinomique avec le principe des médias sociaux. En revanche, le fait de devoir adapter les formes de modération et contrôle aux structures hiérarchiques de l'entreprise nous questionne. Ce principe est cohérent avec des objectifs de formation car il crée des formes de médiation mais l'est moins si l'on considère les principes des médias sociaux qui consistent à aplanir ces formes de hiérarchies pour valoriser la parole de chacun modérée par le collectif. Nous nous demandons s'il est réaliste de faire porter au validateur une charge de travail aussi importante. Son implication est en effet critique pour garantir ce type de fonctionnement. Nous nous demandons de plus si ces exigences sont effectivement pérennes sur le long terme ou si elles ne constituent pas une étape d'acceptation dans le cycle de conception, comme une forme de protection temporaire qui devrait tomber une fois l'usage de ce type de plateforme installée dans l'entreprise.

Une évaluation a été réalisée après 3 mois d'utilisation et montre des résultats prometteurs [35]. Sur cette base, nos prochains objectifs consisteront à étendre le déploiement de la plateforme à l'ensemble des CO pour observer la recevabilité des principes à l'ensemble de l'organisation, les effets d'apprentissage informel et répondre aux questionnements plus généraux sur les formes de modération. Plus globalement, il serait intéressant de mener d'autres expérimentations pour identifier les éléments reproductibles de la méthode de conception, par exemple concernant les séquences et durées des cycles ou le nombre de participants et focus groups optimaux, pour proposer une méthodologie plus opérationnelle.

## Bibliographie